\documentclass[twocolumn,dvipdfmx]{aastex6}
\usepackage{color} 

 \newcommand{\eco}{${\rm C^{18}O}\,$} 	
 \newcommand{\1}{($J=1-0$)} 

\shorttitle{TMC-1}
\shortauthors{Dobashi et al.}

\begin{document}


\title{Discovery of CCS Velocity-Coherent Substructures in the Taurus Molecular Cloud 1}


\author{
Kazuhito Dobashi\altaffilmark{1},Tomomi Shimoikura\altaffilmark{1,2}, Tetsu Ochiai\altaffilmark{1}, \\
Fumitaka Nakamura\altaffilmark{3,4}, Seiji Kameno \altaffilmark{5},
Izumi Mizuno \altaffilmark{6,7,8}, and Kotomi Taniguchi \altaffilmark{4,6,9,10}
}

\affil{\scriptsize{\rm $^1$ dobashi@u-gakugei.ac.jp}}
\altaffiltext{1}{Department of Astronomy and Earth Sciences, Tokyo Gakugei University, Koganei, Tokyo  184-8501, Japan}
\altaffiltext{2}{Otsuma Women's University, Chiyoda-ku, Tokyo, 102-8357, Japan}
\altaffiltext{3}{National Astronomical Observatory of Japan, Mitaka, Tokyo 181-8588, Japan}
\altaffiltext{4}{The Graduate University for Advanced Studies (SOKENDAI), 2-21-1 Osawa, Mitaka, Tokyo 181-0015, Japan}
\altaffiltext{5}{Joint ALMA Observatory, Alonso de C\'{o}rdova 3107 Vitacura, Santiago, Chilie}
\altaffiltext{6}{Nobeyama Radio Observatory, National Astronomical Observatory of Japan 462-2 Nobeyama, Minamisaku, Nagano 384-1305, Japan}
\altaffiltext{7}{Department of Physics, Faculty of Science, Kagoshima University, 1-21-35 Korimoto, Kagoshima, Kagoshima 890-0065, Japan}
\altaffiltext{8}{East Asian Observatory, 660 N. A'oh\={o}k\={u} Place, University Park, Hilo, HI 96720, USA}
\altaffiltext{9}{Virginia Initiative on Cosmic Origins Fellow}
\altaffiltext{10}{Departments of Astronomy and Chemistry, University of Virginia, Charlottesville, VA 22904, USA}


\begin{abstract}
We present the results of mapping observations toward a nearby starless filamentary cloud,
the Taurus Molecular Cloud 1 (TMC-1), in the CCS($J_N=4_3-3_2$, 45.379033 GHz) emission line,
using the Nobeyama 45-m telescope. 
The map shows that the TMC-1 filament has a diameter of $\sim 0.1$ pc and a length of $\sim$ 0.5 pc
at a distance of 140 pc.
The position-velocity diagrams of CCS clearly indicate the existence of velocity-coherent 
substructures in the filament.   
We identify 21 substructures that are coherent in the 
position-position-velocity space by eye.
Most of the substructures are elongated along the major axis of the TMC-1 filament.
The line densities of the subfilaments are close to the critical line density for the equilibrium
($\sim17$ $M_\sun$ pc$^{-1}$ for the excitation temperature of $10$ K), suggesting
that self-gravity should play an important role in the dynamics of the subfilaments.
\end{abstract}


\section{INTRODUCTION} \label{sec:intro}

The {\it Herschel} observations have revealed that filamentary structures are ubiquitous in molecular clouds.  
\citet{andre14} found that 70 \% of prestellar cores in Aquila Rift are distributed along the dense filaments with $A_V \gtrsim 7$ mag, 
suggesting that the filaments play a dominant role in the formation 
of prestellar cores and thus star formation process.  According to  \citet{andre14}, 
the filaments discovered by the {\it Herschel} observations appear to have a typical diameter of 0.1 pc. 
The dense filaments are also supercritical for radial collapse, and thus gravitational fragmentation
is likely to lead to formation of self-gravitating dense cores where stars are created.
However, the origin of the 0.1 pc supercritical filaments remains uncertain. 

One way to understand  its origin is to investigate the kinematics and the internal structure 
of the filaments using molecular line emission which can distinguish the structures 
overlapping along the line of sight.
On the basis of \eco \1 and N$_2$H$^+$ \1 data obtained by the FCRAO
14-m  telescope, \citet{Hacar2013} analyze the internal structure of a long filament in L1495/B213
in the Taurus molecular cloud.
They searched for components having similar
line-of-sight velocities in neighboring pixels using an algorithm named Friend in VElocity (FIVE),
and found that the filament identified by the {\it Hershel} observations contains velocity-coherent
elongated structures which cannot be distinguished in the dust continuum emission map. 
They call the substructures ``fibers".  The typical length of the fibers is about 0.5pc, and
their typical line masses are estimated to be a few times 10 $M_\odot$ pc$^{-1}$,
which is comparable to the critical line mass for gravitational contraction
with the temperature of $\sim$ 10 K.  
Similar substructures are also found in other filaments \citep[e.g.,][]{Hacar2017}.
However, it remains uncertain how ubiquitous the fibers are in molecular cloud.
Their physical origin is also to be elucidated.
To better understand the filamentary structure in molecular clouds, 
we search for velocity coherent structures similar to those probed by \citet{Hacar2013} 
toward the other filament in Taurus known as the Taurus Molecular Cloud 1 (TMC-1).

TMC-1 is a starless dense filamentary cloud and is
very abundant in carbon-chain molecules such as CCS and HC$_3$N
\citep[e.g.,][]{Hirahara1992,Suzuki1992,Taniguchi2016a,TaniguchiSaito2017}.
These molecules are known to appear in an early stage of cloud evolution,
and have been detected in other low mass clouds \citep[e.g.,][]{Hirota2009,Shimoikura2012,Taniguchi2017a}
as well as in massive clouds forming high mass stars
\citep[e.g.,][]{Taniguchi2016b,Taniguchi2017b,Taniguchi2018a,Taniguchi2018b,Taniguchi2018c}
and star clusters \citep[e.g.,][]{Shimoikura2015,Shimoikura2018a},
but they are absent in more evolved clouds such as W40 \citep[e.g.,][]{Shimoikura2015,Shimoikura2018b}.
The high abundance of CCS and HC$_3$N in TMC-1
indicates that this region is young in terms of chemical composition.
A chemical evolution model \citep{Suzuki1992, Ruaud2016} infers an age of
TMC-1 of $\sim$ $10^5$ years.

In this paper, we use the CCS($J_N=4_3-3_2$) emission line to investigate the internal structure.
The typical CCS spectral line profile in TMC-1 can be fitted by
multi-Gaussian profiles,  suggesting the existence of substructures.
Recently, we discovered that the CCS line profile 
at the cyanopolyyne peak in TMC-1 consists of four Gaussian components 
with different line-of-sight velocities \citep{Dobashi2018}.
We summarize the main results of our previous work in Appendix \ref{sec:review} of this paper.
The existence of the multiple velocity components implies
that TMC-1 contains smaller substructures previously found in other regions.

In fact, \citet{Feher2016} found four subfilaments in TMC-1 in NH$_3$ with
the Effelsberg 100-m telescope. Their map has an effective angular resolution of 
$\sim 50\arcsec$ (more accurately, 40$\arcsec$ grids with the HPBW of 37$\arcsec$
in position switching mode).
In an earlier study, \citet{Langer1995} already reported that TMC-1 contains many small blobs in CCS
and these small blobs seem to be gravitationally unbound, and the typical mass is
estimated to be only $\sim0.1$ $M_\odot$.  Substellar-mass self-gravitating structures are also found in the 
other star-forming region, $\rho$ Ophiuchus B2 clump \citep{Nakamura2012}. 
These previous studies suggest the existence of substructures 
in dense structures such as dense cores.

In this paper, we identify the velocity-coherent substructures in TMC-1 in
the CCS ($J_N=4_3-3_2$) emission line using the Z45 receiver installed on the Nobeyama 45-m telescope.
For the physical conditions of TMC-1, the $J_N=4_3-3_2$ transition line is the brightest
among the rotational transition lines of CCS \citep[see e.g.,][]{Suzuki1992,Wolkovitch97}.
Taking into account the chemical age of TMC-1, 
the CCS line should have more advantage than other molecular lines
to search for the internal structure of this region.
In Section \ref{sec:obs}, we describe the CCS mapping observations.
In Section \ref{sec:distributions}, we demonstrate the global CCS distributions in TMC-1.
In Section \ref{sec:substructure}, we identify velocity-coherent structures in
the position-position-velocity space. We have identified 21 structures by eye.
In Section \ref{sec:mass}, we derive the masses of the substructures.
Finally, we briefly discuss the dynamics of the TMC-1 filament
in Section \ref{sec:dis}, and summarize our conclusions in Section \ref{sec:conclusions}.


\section{OBSERVATIONS} \label{sec:obs}

We carried out CCS($J_N = 4_3-3_2$, 45.379033 GHz) and
HC$_3$N($J = 5-4$, 45.490316 GHz) mapping observations in the On-The-Fly (OTF) mode
toward TMC-1 in 2015 May with the Nobeyama 45-m telescope.
For details of the observations, see \citet{Dobashi2018}.
In this paper, we use only CCS line data.
We used the dual-polarization Z45 receiver at 45 GHz band
\citep{nakamura15}.
The main beam efficiency and half power beam width (HPBW) 
were $\eta= 0.7$  and $\Theta \simeq 37\arcsec$, respectively. 
At the back end, we used the 4 sets of a 4096 channel SAM45 digital spectrometer 
with a frequency resolution of 3.81 kHz, corresponding to 
$\sim0.025$ km s$^{-1}$ at 45 GHz.
The typical system temperatures during the observations were 150 K.
The telescope pointing was checked every 1 hr by observing a SiO maser
source, NML Tau, and the typical pointing offset was better
than $3\arcsec$ during the observations. By applying a
convolution scheme with a spheroidal function, we obtained the final map
having an effective angular resolution of $\sim49\arcsec$.

\begin{figure}
\begin{center}
\includegraphics[scale=0.9]{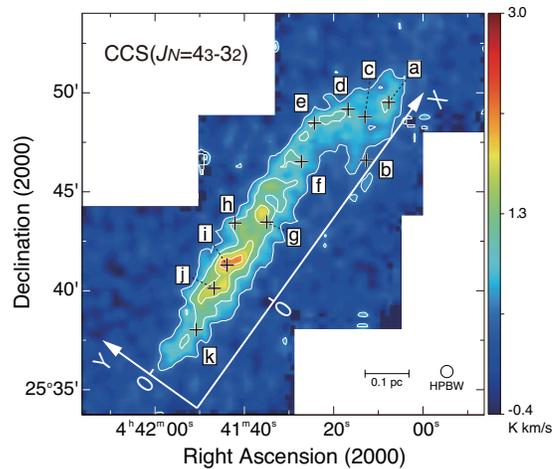}
\caption{The CCS ($J_N=4_3-3_2$) intensity map integrated from 5 km s$^{-1}$ to 7 km s$^{-1}$.
We set $X$ and $Y$ axes as shown in the panel. The position of the CCS peak
is located at $(X, Y)=(0\arcmin, 0\arcmin)$
whose equatorial coordinates are
$\alpha_{\rm J2000}$=04$^{\rm h}$41$^{\rm m}$43.87$^{\rm s}$ and 
$\delta_{\rm J2000}$=+25$^{\circ}$41$\arcmin$17.7$\arcsec$.
Plus signs labeled ``a"--``k" indicate the positions where the CCS spectra shown in
Figure \ref{fig:fig2} are sampled. Positions with labels ``b", ``c", and ``i"  correspond to IRAS 04381+2540,
the NH$_3$ peak position \citep{Feher2016}, and the CCS peak position, respectively. 
\label{fig:fig1}}
\end{center}
\end{figure}

\begin{figure}
\begin{center}
\includegraphics[scale=0.5]{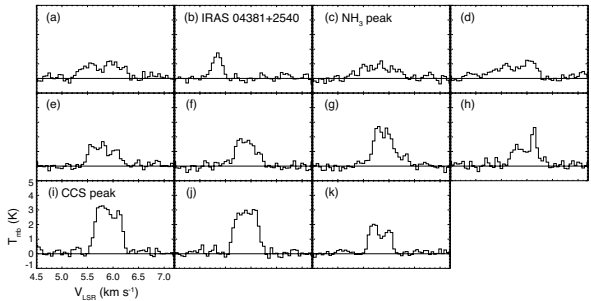}
\caption{
Panels (a)--(k) show the CCS spectra sampled at the positions ``a"--``k"
in Figure \ref{fig:fig1} in this order.
Spectra in panels (b), (c), and (i) are observed toward IRAS 04381+2540, the NH$_3$ peak \citep{Feher2016},
and the CCS peak, respectively. 
All of the spectra are smoothed to the 0.05 km s$^{-1}$ velocity resolution.
\label{fig:fig2}}
\end{center}
\end{figure}


\section{Distributions of the CCS Emission}\label{sec:distributions}

In Figure \ref{fig:fig1}, we show the intensity map of the CCS 
($J_N=4_3-3_2$) emission integrated from 5 km s$^{-1}$ to 7 km s$^{-1}$.
The TMC-1 filament has an aspect ratio of $\sim 7$ with a major axis of $\sim 20\arcmin$ and minor axes of $\sim 3 \arcmin$.  
At a distance of 140 pc \citep[e.g.,][]{Elias1978}, the minor axis has a length of about 0.1 pc, comparable to
the typical diameter of the {\it Herschel} filament. 
The TMC-1 filament detected in the CCS intensity map appears to be
consistent with that of the {\it Herschel} image \citep{Feher2016}.
Hereafter, we define the $X$ and $Y$ axes of the filament as indicated in Figure \ref{fig:fig1}.

The CCS emission is stronger in the southern part of the filament which appears to fragment 
along the major axis.
The position of the strongest emission,
$\alpha_{\rm J2000}$=04$^{\rm h}$41$^{\rm m}$43.87$^{\rm s}$ and 
$\delta_{\rm J2000}$=+25$^{\circ}$41$\arcmin$17.7$\arcsec$,
is designated by
the plus sign labeled ``i" in Figure \ref{fig:fig1}.
This position is often referred to as cyanopolyyne peak (CP).
We found that the CP position is slightly shifted (by a few tens arcsecs)
from the originally-reported position
by earlier studies \citep[e.g., ][]{Hirahara1992}.  We believe that this discrepancy comes from the different observation mode.
We conducted OTF observations to obtain the map.  On the other hand, previous observations
were done by simple position-switch mode.  Taking into account this difference, we believe that 
the position we identified should be more accurate.

\begin{figure}
\begin{center}
\includegraphics[scale=0.9]{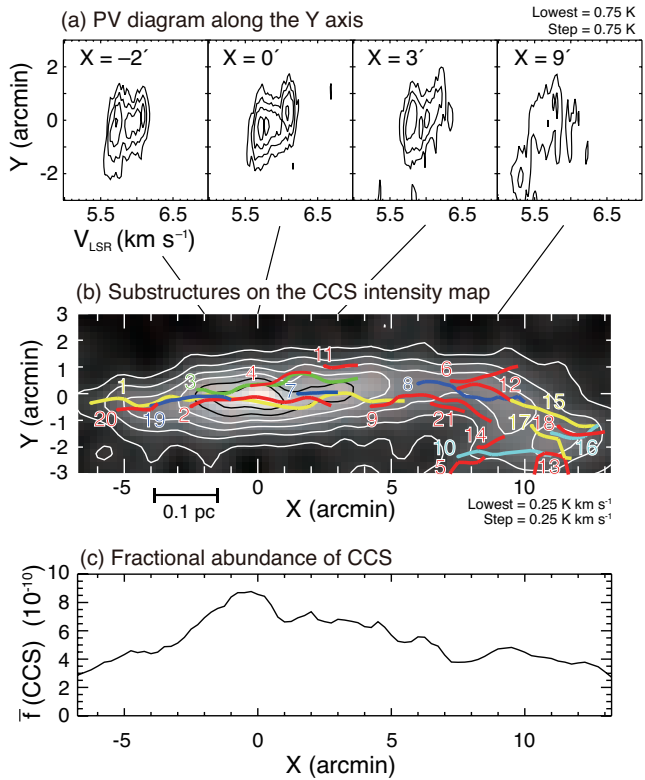}
\caption{
(a) Position-velocity (PV) diagram measured along the $Y$ axis at
$X=-2\arcmin$, $0\arcmin$, $3\arcmin$, and $4\arcmin$, where $X$ and $Y$ axes are drawn in Figure \ref{fig:fig1}.
(b) Distributions of the subfilaments (labelled 1--21 ) shown on the CCS intensity map.
(c) Fractional abundance of CCS measured as a function of $X$.
\label{fig:fig3}}
\end{center}
\end{figure}

In Figure \ref{fig:fig2}, we show examples of the CCS spectra observed toward positions labeled ``a"--``k"
in Figure \ref{fig:fig1}. \citet{Feher2016} reported that in TMC-1 there are two velocity components at most observable in NH$_3$, but the CCS emission line appears complex consisting of more velocity components
over the filament.
In Figure \ref{fig:fig3}(a), we present the position-velocity diagrams of the CCS emission 
along four cuts parallel to the $Y$ axis. 
The diagrams also imply the existence of multiple components overlapping
along the line of sight.
The existence of the substructures was first pointed out by \citet{Langer1995}
based on the different transition data of CCS.
\citet{Peng1998} observed an $8\arcmin \times 8\arcmin$ region around the CP position
in the CCS($J_N=2_1-0_1$) and CCS($J_N=4_3-3_2$) emission lines with similar angular and
velocity resolutions to our observations. The CCS($J_N=4_3-3_2$) spectra they obtained
are very similar to our data in terms of the shapes, peak temperatures, and line widths. They 
divided the emission into three filaments by fitting the spectra with three simple Gaussian functions
to find clumps along the filaments. In Appendix \ref{sec:comparison}, we compare our results with these
earlier studies \citep{Peng1998, Feher2016}.

Recently, we conducted a detailed analysis of the CCS and HC$_3$N line profiles at
the CP position, taking into account the effects of radiative transfer \citep{Dobashi2018}.
We found that the line profile consists of four Gaussian components 
with different line-of-sight velocities, which was confirmed by analyzing optically thin
($\tau \lesssim 0.17$) hyperfine lines of HC$_3$N.
We used data having an extremely fine velocity resolution
of 0.4 m s$^{-1}$ obtained with a spectrometer named PolariS \citep{Mizuno2014}, 
and estimated the centroid velocity of each component very accurately.
Main results of our previous work related to this paper are summarized
in Appendix \ref{sec:review}. For details, see \citet{Dobashi2018}.
The existence of the multiple components
is in good agreement with our position-velocity map
with a coarser velocity resolution.

\begin{figure}
\begin{center}
\includegraphics[scale=0.9]{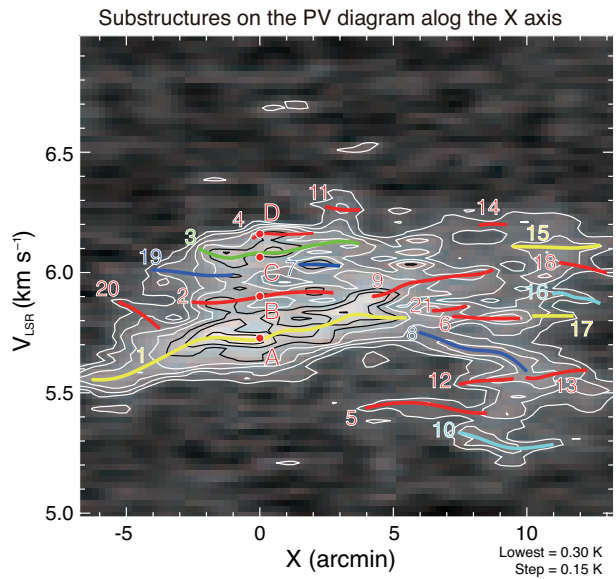}
\caption{
Distributions of the subfilaments shown on the PV diagram along the $X$ axis.
The PV diagram is produced by averaging the CCS spectra over the $Y$ range $-3\arcmin<Y<3\arcmin$
at every $X$. Red circles labelled A--D at $X=0\arcmin$ denote the centroid velocities of the four components
identified by \citet{Dobashi2018}.
\label{fig:fig4}}
\end{center}
\end{figure}


\section{Velocity-Coherent Substructures}\label{sec:substructure}

As shown in
Figure \ref{fig:fig2} and
Figure \ref{fig:fig3}(a), the CCS spectra at several positions have
multiple components along the line of sight.
In the following, we identify the possible substructures in TMC-1
based on our CCS data.
\citet{Hacar2013} first fitted the C$^{18}$O$(J=1-0)$ spectra  in L1495/B213
with $N$-component Gaussian functions and 
searched for velocity components having
similar radial velocities in neighboring pixels using the FIVE algorithm.
Following  \citet{Hacar2013}, we identify velocity-coherent structures
in position-position-velocity space.
In the case of the CCS data obtained in this work, however,
it is not always easy to identify such substructures
with an automatic method such as the FIVE or ``clumpfind" \citep{Williams1994} algorithms,
because the emission line is not optically thin in the dense region ($\tau=1-2$)
and two or more components with a small velocity difference ($0.1-0.2$ km s$^{-1}$)
are aligned on the same line of sight
(e.g, see Appendix \ref{sec:review}),
which makes it difficult to separate them reliably.
We therefore attempt to identify such structures in the position-position-velocity space by visual inspection for simplicity.

Our procedure to find the subfilaments is as follows; as shown in Figure \ref{fig:fig1},
we first set the $X$ and $Y$ axes parallel and orthogonal to the elongation of TMC-1, respectively,
and resampled the spectra at the 15$\arcsec$ grid along the $ Y$axis
by convolving the original CCS spectra with a 2 dimensional Gaussian function (FWHM$=30\arcsec$),
and averaged the resulting spectra in the range $-3\arcmin<Y<3\arcmin$.
We calculated the averaged spectra
at every $X$ grid to produce the position-velocity (PV) diagram along
the $X$ axis.
Resulting PV diagram is shown in Figure \ref{fig:fig4}.
For a single CCS spectrum (e.g., those in Figure \ref{fig:fig2}), it is not always
easy to identify the individual velocity components.
The PV diagram providing information of neighboring spectra
is useful to find candidates of the velocity-coherent substructures
which often appear as a ridge running parallel to the $X$ axis in the diagram.
To investigate distributions of the substructures on the $XY$ plane,
we selected a position $(X,V_{\rm LSR})$ along the ridge in the PV diagram,
and calculated the CCS intensity for all of the $Y$ positions
at the selected $(X,V_{\rm LSR})$ position, and plotted the peak intensity
position on the $XY$ plane. We repeated this procedure for all of the positions
along the noticeable ridges in the PV diagram, and regarded continuous 
peaks both in the PV diagram and $XY$ plane as a substructure.

In total, we found 21 CCS substructures.
Their spatial and velocity distributions are delineated by the solid lines with different colors in Figures \ref{fig:fig3}(b) and \ref{fig:fig4}.
The positions and physical quantities such as lengths, widths, and masses 
of the identified substructures are summarized in Table \ref{tab:tab1}.
The length ranges from $\sim 0.1$ pc to $\sim 0.5$ pc with a width of $0.05 - 0.1$ pc.
Thus, the aspect ratios (the length/width ratios) are estimated to be about 0.3$-$5.7.
Most of the substructures identified have aspect ratios greater than unity, indicating that
they are filamentary.  The substructures with aspect ratios
smaller than unity are blobs or cores, instead of the filaments.

As we describe in Appendix \ref{sec:uncertainties}, there may be rather large ambiguities in
our analyses which mainly arise from our method to find the substructures by visual inspection.
However, most of the major substructures in TMC-1 should be detected in this survey,
which provides us with important information on the structure of TMC-1.
First,
we found four velocity components toward the CP position, and named them
A, B, C, and D in the order of their centroid velocities.
In Figure \ref{fig:fig4}, we indicate their locations in the PV diagram.
The components apparently correspond to the substructures labeled ``1"--``4",
suggesting that the velocity components A--D observed toward the CP position
by our previous study
are likely to be a part of the distinct substructures.

Second, we found that there is a systematic velocity gradient orthogonal to 
the elongation of TMC-1, which
is evident in the PV diagrams in Figure \ref{fig:fig3}(a).
In addition, the identified substructures in the southern part of TMC-1
at $-5\arcmin \lesssim X \lesssim 3\arcmin$ in Figure \ref{fig:fig3}(b)
are well-aligned and concentrated around the axis of the main filament (around $Y\simeq 0\arcmin$),
while those in the northern part ($5\arcmin \lesssim X \lesssim 13\arcmin$)
are more widely and randomly distributed.
The systematic velocity gradient seen in the southern part may infer the global infall motion
in the main filament.
In fact, on the basis of the detailed radiative transfer calculations with CCS and HC$_3$N line profiles,
we proposed in our previous study \citep{Dobashi2018}
that the components A, B, C, and D having lower radial velocities
in this order are aligned along the line of sight from farther side in this order,
and concluded that the four components are shrinking, getting closer to one another as a whole.


\begin{deluxetable*}{cccccccccccc} 
\tablecaption{Parameters of the velocity-coherent components
\label{tab:tab1}} 
\tabletypesize{\scriptsize}
\tablehead{ 
\colhead{} & \multicolumn{5}{c}{Peak Position} &  \colhead{} & \colhead{} & \colhead{} & \colhead{} & \colhead{} & \colhead{} \\
\cline{2-6}
 \colhead{No.} & \colhead{R.A.(J2000)}   & \colhead{Dec.(J2000)}  & \colhead{$T_{\rm mb,max}$}  & \colhead{$V_{\rm LSR}$}  & \colhead{$N({\rm CCS})$} &  \colhead{Length} & \colhead{${\rm Width}$} & \colhead{${\rm Aspect}$} & \colhead{Flux}& \colhead{Mass}& \colhead{Line density} \\
 \colhead{}     &\colhead{}                        & \colhead{}                     & \colhead{(K)}                   & \colhead{(km s$^{-1}$)}    &  \colhead{(10$^{12}$ cm$^{-2}$) }  &  \colhead{(pc)}        &  \colhead{(pc)} & \colhead{${\rm Ratio}$} &  \colhead{(K kms$^{-1}$ arcmin$^2$)}  &  \colhead{($M_\odot$)} & \colhead{($M_\odot$ pc$^{-1}$)}
}
\startdata 
 1 & 4$^{\rm h}$41$^{\rm m}$43.0$^{\rm s}$ & 25$^{\circ}$41$\arcmin$ 9$\arcsec$ & 3.37 & 5.72 & 9.65 & 0.498 & 0.088 &  5.655 & 1.84 & 11.08 & 22.25 \\
 2 & 4$^{\rm h}$41$^{\rm m}$44.9$^{\rm s}$ & 25$^{\circ}$40$\arcmin$32$\arcsec$ & 2.96 & 5.89 & 8.47 & 0.220 & 0.073 &  3.034 & 0.71 &  3.57 & 16.18 \\
 3 & 4$^{\rm h}$41$^{\rm m}$46.1$^{\rm s}$ & 25$^{\circ}$41$\arcmin$ 2$\arcsec$ & 3.13 & 6.07 & 8.97 & 0.257 & 0.075 &  3.448 & 0.79 &  3.96 & 15.41 \\
 4 & 4$^{\rm h}$41$^{\rm m}$44.8$^{\rm s}$ & 25$^{\circ}$41$\arcmin$26$\arcsec$ & 2.68 & 6.16 & 7.66 & 0.096 & 0.076 &  1.259 & 0.33 &  1.60 & 16.63 \\
 5 & 4$^{\rm h}$41$^{\rm m}$15.4$^{\rm s}$ & 25$^{\circ}$45$\arcmin$56$\arcsec$ & 1.55 & 5.42 & 4.44 & 0.207 & 0.058 &  3.584 & 0.28 &  2.14 & 10.30 \\
 6 & 4$^{\rm h}$41$^{\rm m}$23.9$^{\rm s}$ & 25$^{\circ}$48$\arcmin$51$\arcsec$ & 1.63 & 5.81 & 4.68 & 0.105 & 0.117 &  0.904 & 0.27 &  2.34 & 22.24 \\
 7 & 4$^{\rm h}$41$^{\rm m}$38.7$^{\rm s}$ & 25$^{\circ}$42$\arcmin$55$\arcsec$ & 2.62 & 6.03 & 7.49 & 0.061 & 0.067 &  0.915 & 0.23 &  1.20 & 19.64 \\
 8 & 4$^{\rm h}$41$^{\rm m}$26.7$^{\rm s}$ & 25$^{\circ}$47$\arcmin$ 8$\arcsec$ & 1.97 & 5.71 & 5.64 & 0.176 & 0.101 &  1.737 & 0.41 &  3.43 & 19.50 \\
 9 & 4$^{\rm h}$41$^{\rm m}$30.4$^{\rm s}$ & 25$^{\circ}$44$\arcmin$40$\arcsec$ & 2.99 & 5.91 & 8.56 & 0.199 & 0.092 &  2.160 & 0.55 &  4.05 & 20.34 \\
10 & 4$^{\rm h}$41$^{\rm m}$15.0$^{\rm s}$ & 25$^{\circ}$46$\arcmin$29$\arcsec$ & 2.05 & 5.32 & 5.86 & 0.146 & 0.071 &  2.046 & 0.26 &  2.27 & 15.57 \\
11 & 4$^{\rm h}$41$^{\rm m}$39.1$^{\rm s}$ & 25$^{\circ}$44$\arcmin$31$\arcsec$ & 1.09 & 6.26 & 3.13 & 0.052 & 0.110 &  0.468 & 0.10 &  0.52 & 10.13 \\
12 & 4$^{\rm h}$41$^{\rm m}$22.8$^{\rm s}$ & 25$^{\circ}$48$\arcmin$21$\arcsec$ & 1.53 & 5.55 & 4.37 & 0.089 & 0.102 &  0.874 & 0.18 &  1.54 & 17.27 \\
13 & 4$^{\rm h}$41$^{\rm m}$ 7.9$^{\rm s}$ & 25$^{\circ}$48$\arcmin$43$\arcsec$ & 1.30 & 5.57 & 3.73 & 0.170 & 0.135 &  1.254 & 0.23 &  2.07 & 12.21 \\
14 & 4$^{\rm h}$41$^{\rm m}$14.5$^{\rm s}$ & 25$^{\circ}$47$\arcmin$56$\arcsec$ & 0.71 & 6.20 & 2.03 & 0.045 & 0.155 &  0.290 & 0.07 &  0.55 & 12.27 \\
15 & 4$^{\rm h}$41$^{\rm m}$12.0$^{\rm s}$ & 25$^{\circ}$49$\arcmin$60$\arcsec$ & 1.38 & 6.11 & 3.94 & 0.140 & 0.095 &  1.483 & 0.27 &  2.52 & 17.95 \\
16 & 4$^{\rm h}$41$^{\rm m}$ 9.3$^{\rm s}$ & 25$^{\circ}$49$\arcmin$34$\arcsec$ & 1.30 & 5.92 & 3.71 & 0.079 & 0.061 &  1.296 & 0.11 &  1.03 & 13.01 \\
17 & 4$^{\rm h}$41$^{\rm m}$12.8$^{\rm s}$ & 25$^{\circ}$48$\arcmin$54$\arcsec$ & 1.01 & 5.82 & 2.89 & 0.088 & 0.141 &  0.625 & 0.11 &  0.96 & 10.94 \\
18 & 4$^{\rm h}$41$^{\rm m}$ 8.0$^{\rm s}$ & 25$^{\circ}$49$\arcmin$58$\arcsec$ & 1.43 & 6.03 & 4.09 & 0.075 & 0.079 &  0.955 & 0.13 &  1.32 & 17.53 \\
19 & 4$^{\rm h}$41$^{\rm m}$48.4$^{\rm s}$ & 25$^{\circ}$39$\arcmin$52$\arcsec$ & 2.66 & 5.99 & 7.60 & 0.125 & 0.064 &  1.940 & 0.32 &  1.91 & 15.36 \\
20 & 4$^{\rm h}$41$^{\rm m}$51.8$^{\rm s}$ & 25$^{\circ}$37$\arcmin$57$\arcsec$ & 1.76 & 5.77 & 5.05 & 0.065 & 0.079 &  0.825 & 0.13 &  1.05 & 16.13 \\
21 & 4$^{\rm h}$41$^{\rm m}$26.1$^{\rm s}$ & 25$^{\circ}$46$\arcmin$26$\arcsec$ & 1.56 & 5.84 & 4.48 & 0.053 & 0.120 &  0.436 & 0.16 &  1.40 & 26.66 \\
\enddata
\tablecomments{Coordinates, $T_{\rm mb,max}$ the maximum brightness temperature, $V_{\rm LSR}$ the centroid velocity,
$N($CCS$)$ the column density are measured at the intensity peak position of substructures. }
\end{deluxetable*}

\section{Masses of the substructures}\label{sec:mass}

To study the dynamical states of the substructures, we derived their masses using the CCS data.
The excitation temperatures of the CCS and HC$_3$N lines are reported to be $T_{\rm ex}\simeq6-12$ K
at the CP position \citep{Dobashi2018}, but it should vary slightly over TMC-1.
For simplicity, we assumed a constant excitation temperature of $T_{\rm ex}=10$ K and
an optically thin case in this paper.
The masses derived in this section only weakly depend on the assumed $T_{\rm ex}$,
and they would vary by $\sim12$ $\%$ at most in the range $T_{\rm ex}=5-15$ K.

First, we estimated the fractional abundance of CCS relative to H$_2$, $f{\rm(CCS)}$.
For the observed positions where the CCS emission is significantly detected with velocity-integrated intensity greater than
$I_{\rm CCS} \geqq 0.2$ K km s$^{-1}$,
we calculated the column density of CCS in a standard way \citep[e.g.,][]{Shimoikura2018a}
at each observed position as
\begin{equation}\label{eq:column_density}
N{\rm (CCS)}=C_0 I_{\rm CCS}  \ 
\end{equation}
where $C_0$ is a function of $T_{\rm ex}$ and is $1.683\times10^{13}$ (K km s$^{-1}$)$^{-1}$ cm$^{-2}$ for $T_{\rm ex}=10$ K.
We then estimated $f{\rm(CCS)}$ as
$f{\rm(CCS)}=N{\rm (CCS)}/N{\rm (H_2)}$
where $N$(H$_2$) is the column density of hydrogen molecules calculated in a standard way \citep[e.g.,][]{Shimoikura2013}
using the C$^{18}$O ($J=1-0$) data available at the Nobeyama 45-m data archive \citep[e.g., see][]{Dobashi2018}.
The fractional abundance of C$^{18}$O is assumed to be $f$(C$^{18}$O$)=1.7\times 10^{-7}$ \citep{Frerking1982}.
Though there is a clear tendency that the derived $f$(CCS) is smaller in the northern part of TMC-1,
it varies rather largely from pixel to pixel.  We therefore smoothed $f$(CCS)
along the $Y$ axis at every $X$ grid to derive the mean CCS fractional abundance $\bar f$(CCS) as a function of $X$,
which we actually used to calculate the molecular masses of the substructures. 
As displayed in Figure \ref{fig:fig3}(c), $\bar f$(CCS)
varies from $2 \times 10^{-10}$ to $9\times 10^{-10}$ within the elongation of TMC-1.

Next, we produced the velocity-integrated CCS intensity map $I_{\rm CCS}^{\rm sub}$ for each substructure.
Assuming that the substructures have a constant radial velocity along the $Y$ axis,
we extracted a value of the brightness temperature $T_{\rm mb}$(CCS) for every $Y$ grid at the velocity of the substructures $V_{\rm LSR}$
measured at each $X$ grid, and we made a map of the brightness temperature of the substructure $T_{\rm mb}^{\rm sub}$($X,Y$).
We further estimated the velocity-integrated intensity maps $I_{\rm CCS}^{\rm sub}$ as $I_{\rm CCS}^{\rm sub}=T_{\rm mb}^{\rm sub}\Delta V^{\rm sub}$
where $\Delta V^{\rm sub}$ is the FWHM line width of the substructures.
The true $\Delta V^{\rm sub}$ should be different from substructure to substructure and should also vary depending on
the positions ($X,Y$). It is however very difficult to quantify  $\Delta V^{\rm sub}$, and we therefore adopted the
mean value $\Delta V^{\rm sub}=0.17$ km s$^{-1}$ of the four velocity components A--D observed at the CP position \citep{Dobashi2018}.
We derived distributions of $N$(CCS) by
substituting  $I_{\rm CCS}^{\rm sub}$ for $I_{\rm CCS}$ in Equation (\ref{eq:column_density}),
and derived the total molecular mass of each substructure by applying $\bar f$(CCS)
and assuming a distance of 140 pc \citep[e.g.,][]{Elias1978}.
The derived masses are summarized in Table \ref{tab:tab1}.
Note that the masses in the table would increase by $\sim10$ $\%$, if we adopt a distance of $\sim148$ pc
reported by \citet{Loinard2007}.
In the table, flux denotes the total $I_{\rm CCS}^{\rm sub}$ summed over the substructures,
and width denotes the mean FWMH width of $I_{\rm CCS}^{\rm sub}$ along the $Y$ axis.
Aspect ratio is the ratio of length to width, and line density is defined to be the mass divided by the length.
The total and mean masses of the 21 substructures in Table \ref{tab:tab1} are
$\sim50$ $M_\odot$ and $\sim2.4$ $M_\odot$, respectively.
Total molecular mass of the entire TMC-1 filament measured in C$^{18}$O is $\sim200$ $M_\odot$
\citep[see Figure 8b of ][]{Dobashi2018}, and thus $\sim25$ $\%$ of the molecular mass is
contained in the substructures.

The dynamical state of an isothermal filamentary cloud is determined by
the line density, which is given as
\begin{equation}\label{mass_density}
m = {2 c_s^2 \over G} \simeq 464 \left({c_s \over  {\rm km \ s^{-1}}} \right)^2 M_\odot {\rm pc^{-1}}   \ ,
\end{equation}
where $c_s$ and $G$ are the isothernal sound speed and gravitational constant, respectively
\citep{Stodolkiewicz1963,Ostriker1964}.
The dust temperature in TMC-1 is measured to be $\sim$ 11 K on the basis of the {\it Herschel} observations \citep{Feher2016}. 
If we adopt the isothermal sound speed at $T\simeq 10$ K ($c_s= 0.2$ km s$^{-1}$), 
the critical line density for the dynamical equilibrium is estimated to be 17 $M_\odot$ pc$^{-1}$.
Though our mass estimates based on the CCS data are rather coarse,
most of the identified substructures have masses comparable to the critical value,
and thus we believe that they should be close to the dynamical equilibrium.  
These nearly-equilibrium structures seem to have similar dynamical properties 
of the fibers identified by \citet{Hacar2017}.
It is also interesting to note that \citet{Peng1998} identified $45$ small clumps in CCS along the filamentary
structures around the CP position, and
they suggested that most of the clumps are gravitationally stable or unbound,
though the optical depth of the emission line and the related radiative transfer were not taken into account in their analyses.


\section{Discussion}\label{sec:dis}

A bundle of aligned substructures as seen in the southern part of TMC-1 can sometimes be recognized in numerical
simulations to study the formation and evolution of dense filaments, although it remains uncertain whether the velocity-coherent 
substructures we found are the same as those found in the numerical simulations.
For example, in the simulations performed by \citet{Dobashi2014} and \citet{Matsumoto2015},
filaments naturally form in dense clumps mainly by turbulence, and
they often consist of smaller substructures.
Similar results are reported by \citet{Moeckel15}.
(see also \citet{Zamora17} and \citet{Smith2016}).
The substructures are gravitationally stable and scarcely collapse spontaneously by themselves,
but when they happen to cross (or collide against) each other,
they can soon form stars due to the increase of density at the intersections
\citep[e.g., see Figure 19 of ][]{Dobashi2014}.
Similar effects of filament-filament collisions on star formation have been reported also
in other star-forming regions \citep[e.g.,][]{Nakamura2014}.
We imagine that TMC-1 consists of such smaller, gravitationally stable substructures,
and as predicted by the simulations, it would soon form stars when the internal
substructures happen to cross each other.

The origin of the velocity-coherent substructures identified here and their role in the star formation process are still uncertain.
They might be created by local cloud turbulence, as predicted by the numerical simulations.
Recently, we found that the substructures Nos. 1 and 4 are infalling with a speed of $\sim$ 0.4 km s$^{-1}$ 
toward the center \citep{Dobashi2018} at the CP position.  
If the TMC-1 filament is infalling toward the major axis,
such a global infall motion may create local denser parts (e.g., substructures Nos.1 and 4) 
generated by shock.
Thus, there is a possibility that we just see such local peaks temporally created in the main filament.
To fully understand the origin and role of the substructures in star formation process,
further investigation on the kinematics and internal structures of {\it Herschel} filaments and substructures
would be needed.

\section{Conclusions} \label{sec:conclusions}
We have observed the Taurus Molecular Cloud 1 (TMC-1) in the CCS($J_N=4_3-3_2$) emission line
at 45 GHz using the Nobeyama 45-m telescope. We found that TMC-1 has a filamentary morphology
with a diameter of $\sim 0.1$ pc and a length of $\sim0.5$ pc.
We identified 21 velocity-coherent substructures in TMC-1 by eye,
and found that the substructures have a mass close to
the critical values for the gravitational equilibrium.


\acknowledgments
This work was financially supported by Grant-in-Aid for Scientific Research 
(Nos. 17H02863, 17H01118, 26287030, 17K00963)
of Japan Society for the Promotion of Science (JSPS). 
K. T. would like to thank the University of Virginia for providing the funds for her postdoctoral fellowship in the VICO research program.


{\appendix

\section{Four velocity components at the cyanopolyyne peak}\label{sec:review}
Based on sensitive CCS and HC$_3$N spectral data, we recently
identified four velocity components at the CP position, and reported the
results in another article \citep{Dobashi2018}.
In the following, we briefly review the methods and results of our previous work.

We observed the CP position with the CCS($J_N=4_3-3_2$) and HC$_3$N($J=5-4$) lines simultaneously
using the Z45 receiver installed on the Nobeyama 45-m telescope. We used a spectrometer
named PolariS \citep{Mizuno2014} which provided an extremely high frequency resolution of 61 Hz
corresponding to the $\sim0.4$ m s$^{-1}$ velocity resolution. Total integration of $\sim30$ hr
suppressed the rms noise of the spectral data to $\Delta T_{\rm a}^{*}\simeq40$ mK
for the $\sim0.4$ m s$^{-1}$ velocity resolution. 
Resulting spectra related to this work are displayed in Figure \ref{fig:fig5}(a) and (b).

The HC$_3$N($J=5-4$) line consists of five hyperfine structures ($F=5-5$, $4-3$, $5-4$, $6-5$, and $4-4$).
Among these, the $F=4-4$ and $5-5$ lines have intrinsically the same intensity, and are well separated
in frequency from the other hyperfine lines. Therefore they can be averaged directly to reduce noise.
The spectrum in Figure \ref{fig:fig5}(a) is the average spectrum of the $F=4-4$ and $5-5$ lines.
The $F=4-4$ and $5-5$ lines are generally optically thin. In fact, the maximum optical depth at
the CP position is $\tau \lesssim 0.17$ \citep[for details, see][]{Dobashi2018}.
Therefore, under the assumption of the Local Thermodynamic Equilibrium (LTE), the spectrum in
Figure \ref{fig:fig5}(a) can be approximated by a simple Gaussian function with $N$ velocity components.
We fitted the average HC$_3$N spectrum with a Gaussian function with $N$ velocity components
leaving the centroid velocity, peak intensity, and line width of each component as free parameters.
We calculated the reduced $\chi^2$ (hereafter, $\chi^2_r$) of the residual of the best fit for increasing
$N$($=1$, $2$, $3$, and so on).
We found that $\chi^2_r$ rapidly decreases up to $N=4$ ($\chi^2_r=1.0016$),
and also that $\chi^2_r$ doesn't change significantly for larger $N(\geqq 5)$.
This indicates that the spectrum consists of $N=4$ distinct velocity components.
We named the four components A, B, C, and D in the order of increasing centroid velocities.
These components are shown by the
solid lines with different colors in Figure \ref{fig:fig5}(a). The fitted centroid velocities
are given in Figure \ref{fig:fig5}(c).

We assumed that the CCS line consists of the same velocity components
having the same centroid velocities as those found in the HC$_3$N line,
because the CCS and HC$_3$N lines should trace similar density regions  ($n_{\rm H_2}=10^{4-5}$ cm$^{-3}$).
We fitted the observed CCS spectrum with a Gaussian function which consists of $N=4$ velocity components
with the fixed centroid velocities, leaving the line width, excitation temperature, and optical depth
of each component as free parameters.
Because the CCS line is not optically thin, radiative transfer should be taken into account in the fitting process.
We looked for the parameters best fitting the observed CCS spectrum for all of the
permutations ($N!=4!=24$) of the relative positions for the four velocity components
along the line-of-sight, and calculated $\chi^2_r$ of the residual of the best fit for each permutation.
As a result, we found that
the minimum $\chi^2_r$ ($=1.063$) can be found when the components A, B, C, and D are lying
from far side to near side from the observer in this order as illustrated in Figure \ref{fig:fig5}(c).
In Figure \ref{fig:fig5}(b),
contributions of the four components to the observed CCS spectrum are shown by the lines
with different colors. Note that the contribution of each component cannot be expressed by
a simple Gaussian function, not only because they are not optically thin, but also because
they are partially absorbed by the other component(s) in the foreground.
In the same way as for the CCS line, we analyzed the other hyperfine lines
of HC$_3$N (i.e., $F=4-3$, $5-4$, $6-5$) which are blended and optically thicker,
and obtained the same results as for the CCS line \citep[see Figure 6 of][]{Dobashi2018}.

In addition to the above results, we pointed out a possibility that the centroid
velocities and the relative locations of the components A--D along the line-of-sight
illustrated in Figure \ref{fig:fig5}(c) may
represent the global infalling motion of TMC-1.

\begin{figure*}
\begin{center}
\includegraphics[scale=0.9]{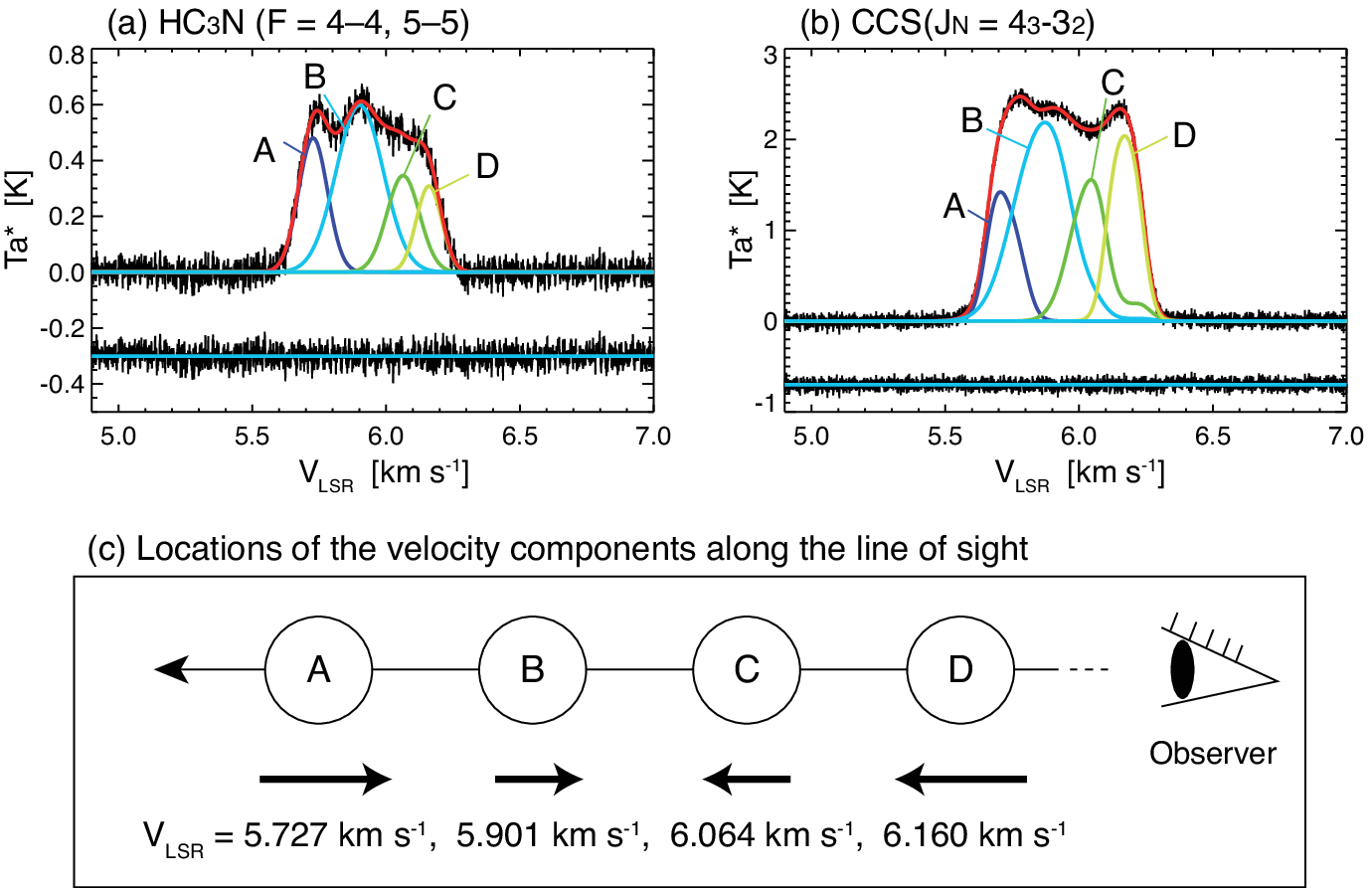}
\caption{
(a) The averaged spectrum of the two hyperfine transitions $F=4-4$ and $5-5$
of the HC$_3$N($J=5-4$) line, and (b) the CCS($J_{N}=4_3-3_2$) line observed
toward the CP peak with the PolariS spectrometer. The original velocity resolution is
0.4 m s$^{-1}$, and it is smoothed to 0.8 m s$^{-1}$ in the panels.
Four velocity components labeled ``A"--``D" identified in the spectra are shown by the
lines with different colors, and their sums best fitting the observed spectra are shown by the red lines. 
Residuals offset by $-0.3$ K and $-0.9$ K in panels (a) and (b), respectively, are also shown.
The spectra are shown in units of the antenna temperature $T_{\rm a}^*$ related to the
brightness temperature as $T_{\rm mb}={\eta}^{-1} T_{\rm a}^{*}$ where $\eta$ ($=0.7$ at 45 GHz)
is the main beam efficiency of the 45-m telescope.
(c) Inferred relative locations of the four velocity components along the line of sight.
The figures are taken from \citet{Dobashi2018}.
\label{fig:fig5}}
\end{center}
\end{figure*}
 
\section{Comparison with earlier studies}\label{sec:comparison}
We compare our results with two earlier studies to investigate the substructures of TMC-1.
One is the CCS observations around the CP position performed by \citet{Peng1998}.
They divided the CCS emission into three filamentary structures
having typical radial velocities of $5.7$, $5.9$, and $6.1$ km s$^{-1}$
by fitting the spectra with a simple Gaussian function with three components at most  (see their Figure 2).
The first two components mainly correspond to our substructures ``1" and ``2", respectively,
and the third component is a mixture of our substructures ``3" and ``4".
A few other substructures identified in our work around the CP position (e.g., ``19" and ``20")
are not recognized or are merged to one of the three components in their work,
because of their assumption of the three components at most over the $8\arcmin \times 8\arcmin$
region they observed.
Except for these points, the results of \citet{Peng1998} are consistent with our results.

The other is the recent work by \citet{Feher2016} who observed the entire TMC-1 filament in NH$_3$. 
In their NH$_3$ data, there can be seen a clear velocity gradient orthogonal to the axis of the main
filament \citep[see Figure 5 of ][]{Feher2016}, which is consistent with what
we see in Figure \ref{fig:fig3}a. They suggested existence of four filamentary substructures
which they call ``F1", ``F2", ``F3", and ``F4".
The lengths ($\sim0.5$ pc) and radius ($0.1-0.2$ pc) of their substructures are
similar to our substructures, but their distributions on the plane of the sky appear very different.
This is because they fitted the spectra only with two velocity components at most
and classified the components using ``$k$-means clustering method".
Their F1--F4 are apparently mixtures of our substructures ``1"--``21",
but it is difficult to identify their correspondence clearly. 
Coincidence of their and our substructures on the plane of the sky
can be summarized as follows;
their F1 contains our substructures ``12", ``15", and ``21", and  F2 contains ``11".
F3 contains ``5", ``10", ``13", and ``14", and F4 contains ``1"--``4", ``7", ``19", and ``20".
Our ``6" is shared by F1 and F2, and ``9" is shared by F2 and F3.
Finally, ``16", ``17", and ``18" are shared by F1 and F3. 

Major differences between our results and those of the earlier studies arise from the
assumptions on the number of velocity components. For example, \citet{Peng1998} and \citet{Feher2016}
assume three and two velocity components around the CP position, while we found four components
toward the position, which agrees with the detailed analyses
in our previous work \citep{Dobashi2018}
as stated in Section \ref{sec:substructure}.

\section{Uncertainties of our analyses}\label{sec:uncertainties}
We describe caveats and ambiguities in our analyses. 
We first describe possible errors arising from our method to find substructures.
We identified 21 substructures which should trace most of the major substructures
in TMC-1. However, there may be some more substructures which escaped detection,
because we searched them by visual inspection.
In addition, our method to find substructures is apparently insufficient to identify and quantify
filaments extending parallel to the $Y$ axis, and there might be
some filamentary substructures extending nearly orthogonal to the $X$ axis.
For example, candidates of such features can be seen at $X\simeq -2\arcmin$ and $X\simeq 6\arcmin$
in Figure \ref{fig:fig3}(b) which we did not analyze in this paper.
In addition, we can identify only one of two (or more) filaments having the same $V_{\rm LSR}$ and
running parallel to the $X$ axis, because the method allows us to pick up only one $Y$ position for a given
($X$,$V_{\rm LSR}$) position. This is also a source of the incomplete sampling.
The lengths of the substructures in Table \ref{tab:tab1} should be the lower limits
to the actual lengths because of this shortcoming of the method.

Finally, we describe errors in our mass estimate of the substructures.
The estimate is basically based on the C$^{18}$O data,
and therefore its accuracy depends on the ambiguity of the assumed fractional abundance of C$^{18}$O.
The assumed fractional abundance ($1.7\times 10^{-7}$)
originally have an ambiguity of a factor of $\sim3$ \citep{Frerking1982}.
In addition, the C$^{18}$O molecule
is known to be adsorbed onto dust in dense regions \citep[e.g.,][]{Bergin2002}.
In fact, the intensity of the C$^{18}$O spectrum at the CP peak is about one half of
what we would expect without adsorption \citep[see Figure 9 of][]{Dobashi2018},
and therefore the derived total mass of the subscrutcures can be underestimated by a factor of $\sim2$ by this effect.
The assumed constant line width ($\Delta V^{\rm sub}=0.17$ km s$^{-1}$) is also a source of
errors, because it is actually varying by a factor of $\sim2$ depending on the velocity components
even at the CP peak \citep[see $\sigma^i$ in Table 3 of][]{Dobashi2018}.
These errors arising from the assumptions on the fractional abundance and the line width
may result in the total ambiguity of a factor of $\sqrt{3^2+2^2+2^2}\simeq4$ in our mass estimate.
There must be some other sources of errors not mentioned in the above, and thus the factor $\sim 4$
should be the minimum estimate of the total uncertainty in the derived mass of the subsctructures.

}



\end{document}